\documentclass[12pt,preprint]{aastex}
\usepackage{graphicx}
\usepackage{amssymb}
\usepackage{amsmath}
\newcommand{\sgr}{SGR 0418+5729}

\begin{document}

\title{Search for the optical counterpart to \sgr}
\author{Martin Durant}
\affil{Department of Astronomy, University of Florida, FL 32611-2055, USA}
\email{martin.durant@astro.ufl.edu}
\author{Oleg Kargaltsev}
\affil{Department of Astronomy, University of Florida, FL 32611-2055, USA}
\author{George G. Pavlov}
\affil{Department of Astronomy and Astrophysics, Pennsylvania State University, PA 16802, USA\\
St.-Petersburg Polytechnical University, Polytechnicheskaya ul. 29, St. Petersburg 195257, Russia}

\begin{abstract}
We report broad-band {\sl Hubble Space Telescope} imaging of the field of soft $\gamma-$ray repeater SGR 0418+5729 with ACS/WFC and WFC3/IR. Observing in two wide filters F606W and F110W, we find no counterpart within the positional error circle derived from {\sl Chandra} observations, to limiting magnitudes $m_{\rm F606W}>28.6$, $m_{\rm F110W}>27.4$ (Vega system), equivalent to reddening-corrected  luminosity limits $L_{\rm F606W}<5\times10^{28}$, $L_{\rm F110W}<6\times10^{28}$\,erg\,s$^{-1}$ for a distance $d=2$\,kpc, at 3-$\sigma$ confidence. This, in turn, imposes lower limits on the contemporaneous  X-ray/optical flux ratio of $\simeq$1100 and X-ray/near-infra-red flux ratio of $\simeq$1000. We  derive an upper limit on the temperature and/or size of any fall-back disk around the magnetar. We also compare the detection limits with observations of other magnetars.
\end{abstract}
\maketitle

\section{Introduction}
Soft Gamma-Ray Repeaters (SGRs) are believed to be {\em magnetars}: neutron stars (NSs)  powered primarily by the decay of a super-strong magnetic field  (Thompson \& Duncan 1995). Their primary observational characteristics are X-ray luminosities higher than the spin-down power, $L_X>\dot{E}$, relatively slow rotation $P\sim1-10$\,s and fast spin-down $\dot{P}\sim10^{-12}-10^{-9}$\,s\,s$^{-1}$, implying characteristic ages $\sim$1000\,yr and  dipole magnetic field strengths  $B_{\rm dipole}\ga10^{13}$\,G. Most SGRs were discovered by their outbursts, either Giant Flares \citep{2005Natur.434.1098H} or rapid burst series \citep{2002Natur.419..142G}. Since the discovery of mid-infrared emission from the magnetar 4U 0142+61 \citep{2006Natur.440..772W}, passive (non-accreting) disks have been suggested to exist around magnetars, and could in principle have some effect on the observed luminosity and/or spin evolution (see \citealt{2009ApJ...702.1309E} and references therein).

So far, only  three magnetars (SGRs and the related Anomalous X-ray Pulsars, AXPs) have  been detected in the optical, SGR 0501+4516 \citep{2008GCN..8229....1O,2008GCN..8160....1F}  4U 0142+61 \citep{2000Natur.408..689H} and 1E 1048.1$-$5937 \citep{2005ApJ...627..376D}. The detections are challenging, due to their faintness and significant reddening (all Galactic magnetars lie close to the Galactic plane). Although other magnetars have been seen in the near-infrared,  SGRs were  detected only during their active states, so their quiescent NIR or optical luminosities are not known. See \citet{2008A&ARv..15..225M} for a review of magnetar properties.

\sgr\ was discovered on 9 June 2009 via magnetar-like bursts detected by the {\sl Fermi} Gamma-Ray Burst Monitor and confirmed with the {\sl Swift} Burst Alert Telescope \citep{2010ApJ...711L...1V}. The 9.1\,s pulsations were first detected by {\sl RXTE} \citep{2009ATel.2076....1G}, while {\sl Swift} XRT spectral fits suggested thermal emission with $kT\approx900$\,eV \citep{2009ATel.2127....1C}. No radio counterpart was found \citep{2009ATel.2096....1L}. \citet{2010MNRAS.405.1787E} summarized the results of the follow-up XRT observations, and found thermal emission slowly cooling, with the emitting area decreasing on a time-scale of months. The extinction column inferred from the X-ray fits was $N_H\approx1.1^{+0.1}_{-0.2}\times10^{21}$\,cm$^{-2}$, lower than for any other Galactic magnetar. Using a limit on the pulsar spin-down rate, $\dot{P}$, \citet{2010Sci...330..944R} derived an upper limit on the pulsar dipole magnetic field, $B_{\rm dipole} < 7.5\times10^{12}$\,G, more typical for an ordinary pulsar than a magnetar. \citet{2011ApJ...732L...4A} argued that this contradicts the magnetar model, and instead supports the fall-back disk model.

\citet{2009ATel.2159....1W} measured the most accurate position of the SGR from {\em Chandra} data, R.A.\ $ = 04^h 18^m 33.867^s,\,$Decl.\ $= +57\degr 32\arcmin 22.91\arcsec$ (J2000), with the positional uncertainty of 0\farcs35 at 95\% confidence. The direction of \sgr\ ($l=147\fdg98$, $b=+05\fdg12$) suggests a distance of 2\,kpc, corresponding to the Perseus Arm \citep{2010ApJ...711L...1V}.
The proximity and low extinction of \sgr\ made it a promising candidate for detailed study in the optical/NIR. \citet{2010MNRAS.405.1787E} failed to detect it in optical imaging with the {\sl Gran Telescopio de Canarias} (GTC), setting a magnitude limit $i' > 25.1$ (SDSS magnitude).
In this paper, we describe our imaging observations of \sgr\ with the {\sl Hubble Space Telescope} (HST). In Section 2 we present the data and  analysis. In Section 3 we discuss the detection limits and conclusions.

\section{Observations and data reduction}
The field of \sgr\ was imaged on 2010 October 19 by {\sl HST} (GO program 12183). In the optical, the Advanced Camera for Surveys, Wide Field Channel (ACS/WFC) was used in conjunction with the broad filter F606W (pivot wavelength 5921\,\AA, RMS bandwidth 672\,\AA; roughly a broader V-band). The 2484\,s, observation consisted of four dithered exposures with the WFC1 aperture. In the near-IR, the Wide Field Camera 3, IR channel (WFC3/IR) was used in MULTIACCUM mode with the broad filter F110W (pivot wavelength 11534\,\AA, RMS bandwidth 4430\,\AA; roughly a broad J-band). The total exposure  was 2811\,s, taken at four dithered pointings with the read parameters SPAR50, NREAD16 (sixteen reads, every 50\,s, giving fifteen net images at each location). The images were processed by the standard pipeline, and combined using {\tt multidrizzle}\footnote{\tt http://stsdas.stsci.edu/multidrizzle/}, which also corrects the images for geometric distortion and performs efficient cosmic ray rejection.

\begin{figure*}
\begin{center}
\includegraphics[width=0.7\hsize,height=0.65\hsize]{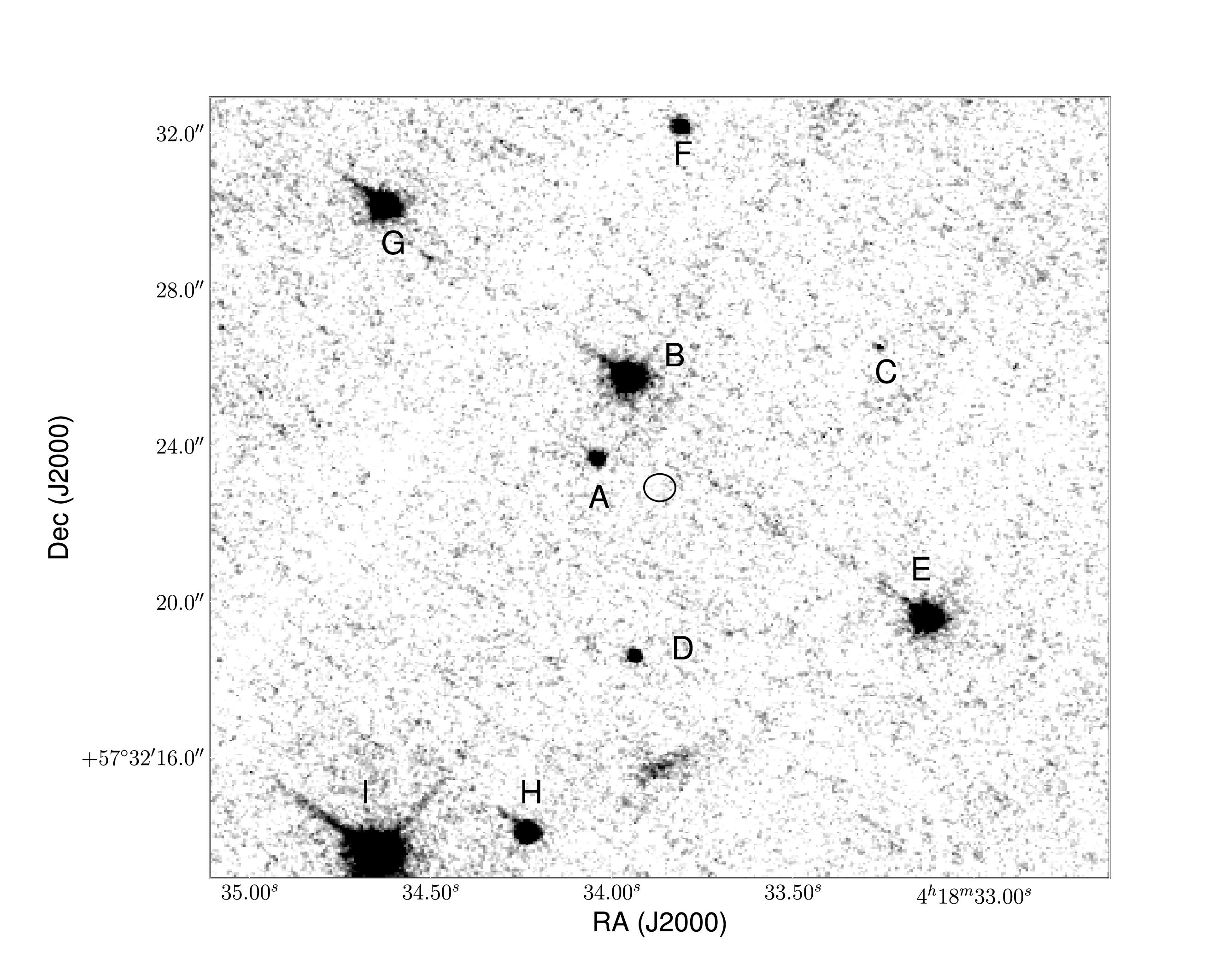}
\includegraphics[width=0.7\hsize,height=0.65\hsize]{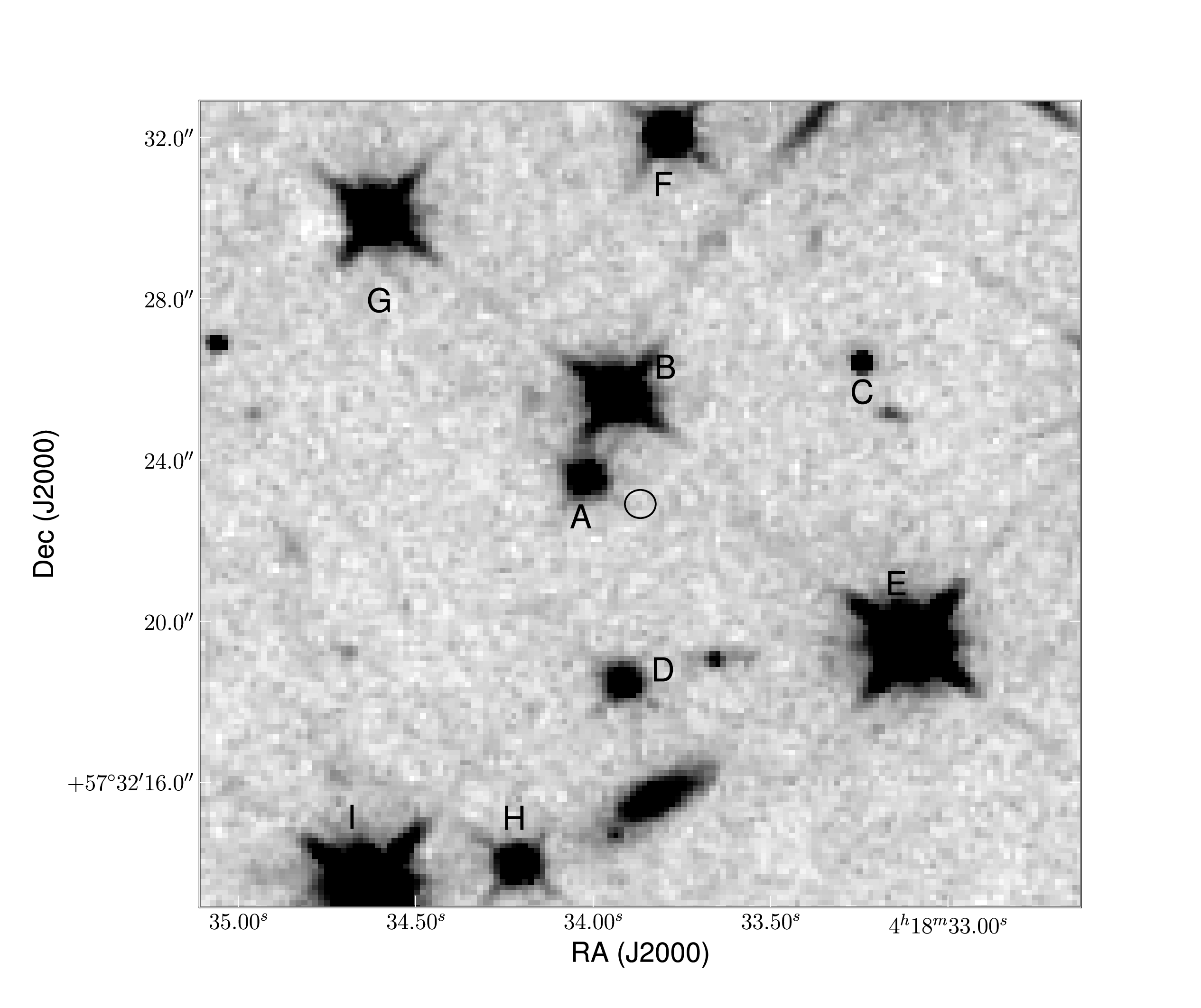}
\caption{Images of the \sgr\ field in the F606W (top) and F110W (bottom) filters. The $r=$0\farcs35 positional error circle from \citet{2009ATel.2159....1W} is indicated. The nearest stars are labelled, and their positions and photometry given in Table \ref{phot}.}\label{im}
\end{center}
\end{figure*}

Photometry was performed using the PSF-fitting package {\tt daophot II} \citep{1987PASP...99..191S}. We tied the astrometry of the F110W image to the 2MASS system \citep{2006AJ....131.1163S} using 22 cross-identified catalog stars and the IRAF task {\tt ccmap}. The astrometric uncertainty was about 0\farcs03 in each coordinate (1-$\sigma$), estimated from the RMS in the residuals. The F606W image was registered with the F110W one using 28 matched stars, with a relative uncertainty of 0\farcs01. Since \citet{2009ATel.2159....1W} also derived the \sgr\ position relative to the 2MASS system, we do not incur any additional systematic error, and hence the overall positional uncertainty is dominated by the X-ray uncertainty, 0\farcs35, at 95\% confidence \citep{2009ATel.2159....1W}. 
The final {\sl HST} images are shown in Figure \ref{im}, together with the positional uncertainty circle. 

We calibrated the photometry using the tabulated zero points for the F110W\footnote{\tt http://www.stsci.edu/hst/wfc3/phot\_zp\_lbn} and F606W\footnote{\tt http://www.stsci.edu/hst/acs/analysis/zeropoints} images, respectively. The uncertainty in the zero points is less than 0.05\,mag.

\section{Results and Conclusions}
\begin{figure*}
\begin{center}
\includegraphics[width=0.7\hsize]{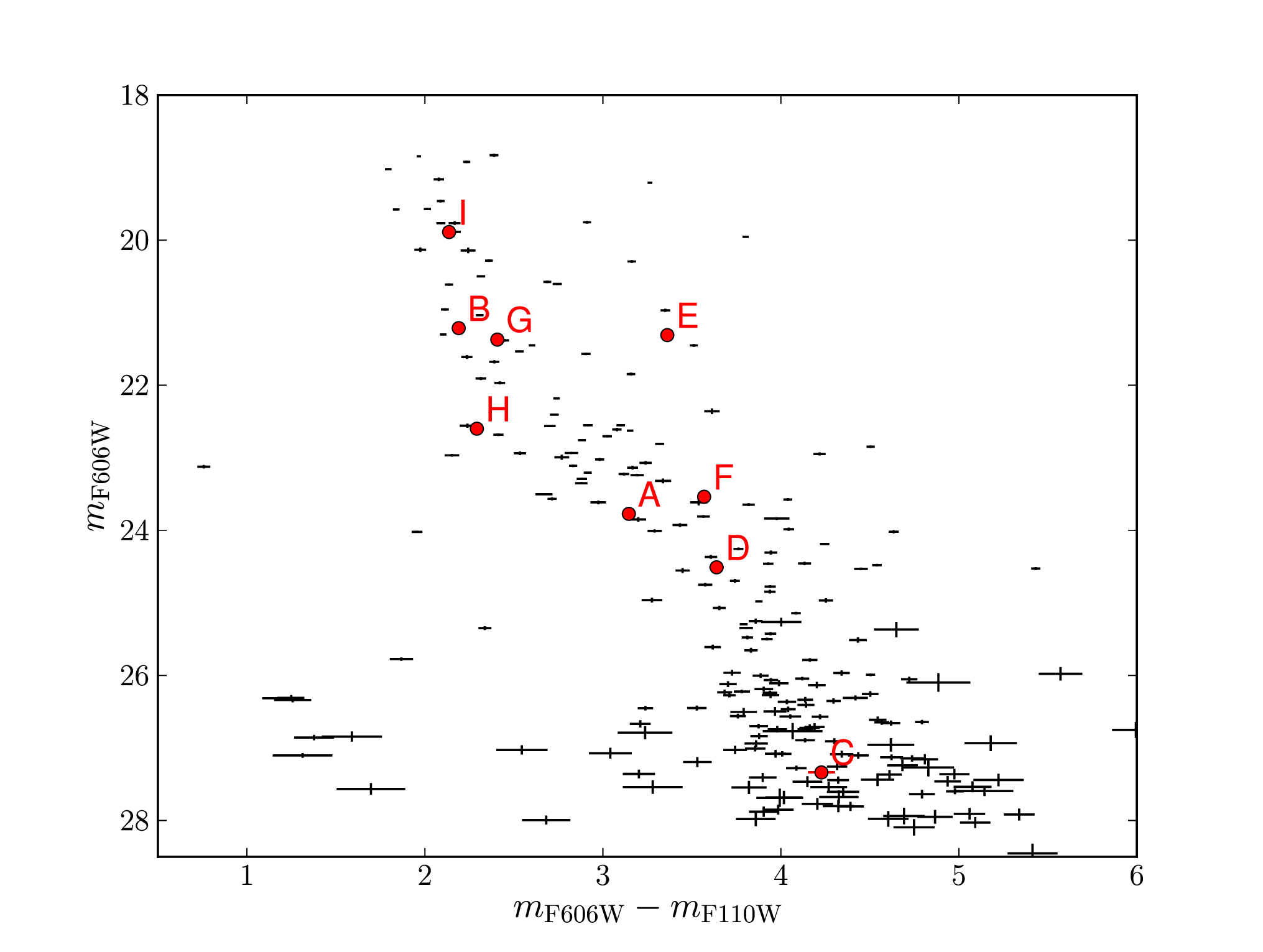}
\caption{Color-magnitude diagram of stars in the field of \sgr. The stars nearest the position of \sgr\ are marked (see Figure \ref{im}).}\label{cmd}
\end{center}
\end{figure*}

In the images shown in Figure \ref{im}, no bright source is seen within or near the SGR error circle. The 3-$\sigma$ limits are $m_{\rm F110W}>27.4$, $m_{\rm F606W}>28.6$ (Vega magnitudes). The locations and magnitudes of the nearest stars, marked in Figure \ref{im}, are listed in Table \ref{phot}. The color-magnitude diagram (CMD) of all the matched field stars is shown in Figure \ref{cmd}, with the point sources from Table \ref{phot} labeled. They are all consistent with main sequence stars behind $A_V\sim1$ reddening, except Star E, which likely is a red giant.

\begin{deluxetable}{ccccc}
\tablecaption{Point sources in the field.\label{phot}}
\tablewidth{0in}
\tablehead{
\colhead{Star} & \colhead{R.A.\ (J2000)} & \colhead{decl.\ (J2000)}  & \colhead{$m_{\rm F606W}$} & \colhead{$m_{\rm F110W}$}
}
\startdata 
A &64.641821 &57.539883 &23.773$\pm$0.013 &20.627$\pm$0.022\\
B &64.641448 &57.540451 &21.213$\pm$0.016 &19.023$\pm$0.020\\
C &64.638586 &57.540671 &27.337$\pm$0.069 &23.109$\pm$0.028\\
D &64.641397 &57.538482 &24.510$\pm$0.029 &20.871$\pm$0.028\\
E &64.638012 &57.538750 &21.308$\pm$0.016 &14.947$\pm$0.017\\
F &64.640854 &57.542242 &23.537$\pm$0.014 &19.968$\pm$0.018\\
G &64.644254 &57.541679 &21.371$\pm$0.015 &18.964$\pm$0.013\\
H &64.642619 &57.537232 &22.559$\pm$0.010 &20.307$\pm$0.024\\
I &64.644363 &57.537078 &19.888$\pm$0.017 &17.752$\pm$0.018
\enddata
\tablecomments{Uncertainties do not include the uncertainty in the photometric zero points. Magnitudes are in the Vega system.}
\end{deluxetable}

\begin{figure*}
\begin{center}
\includegraphics[width=0.95\hsize]{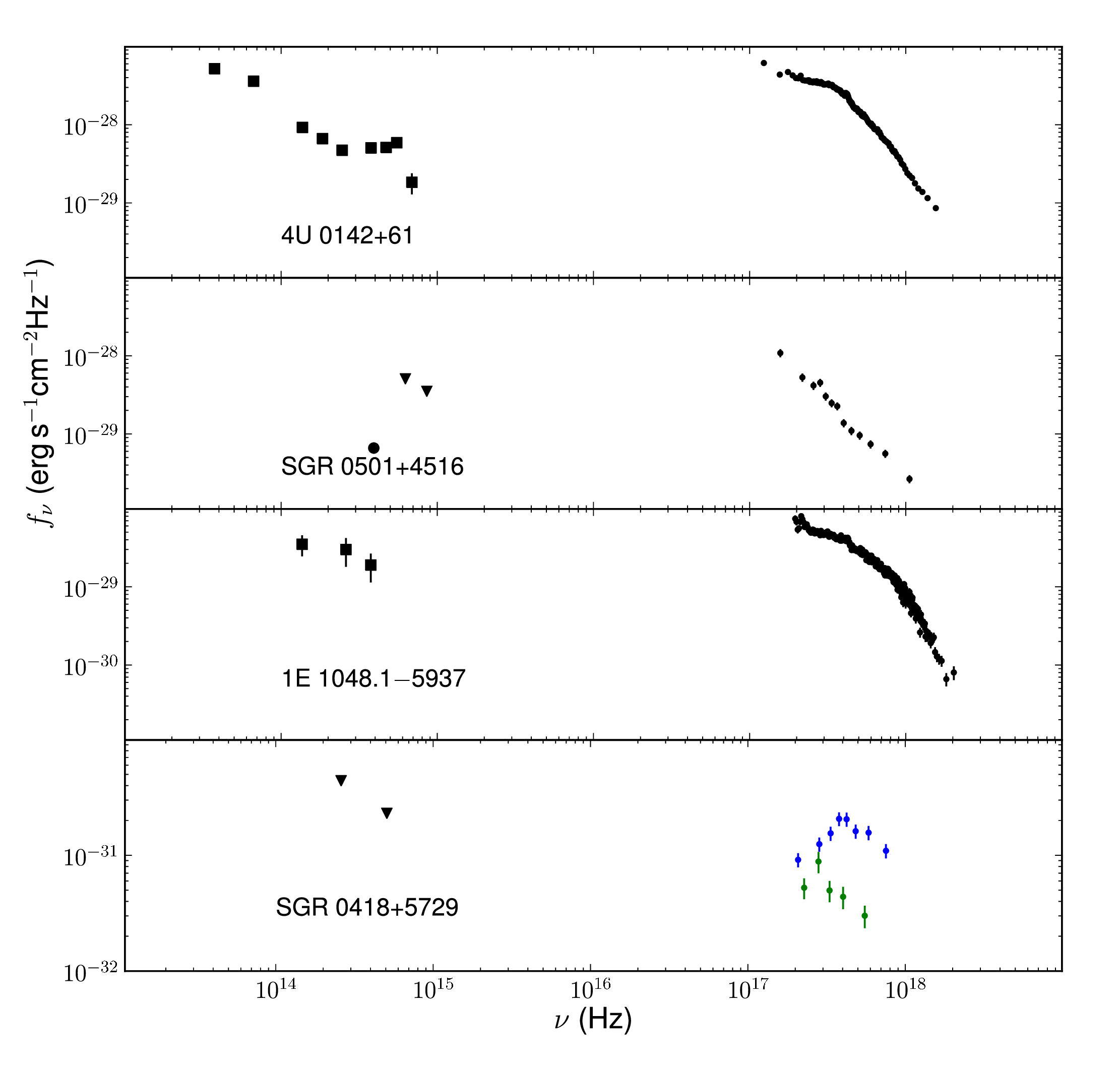}
\caption{Multiwavelength spectra of the optically detected magnetars, together with \sgr. For \sgr, we show two X-ray spectra from 2010 July 23 (upper, blue) and 2010 November 29 (lower, green), and the spectral flux limits established in this paper.}\label{specs}
\end{center}
\end{figure*}

\citet{2010Sci...330..944R} give the  SGR X-ray flux  $F_{\rm X}=(1.2\pm0.1)\times10^{-13}$\,erg\,s$^{-1}$cm$^{-2}$ (0.5--10\,keV, absorbed; {\sl Chandra} ObsID 12312) on 2010 July 23, which is a factor of 150 fainter than during the
 discovery observations a year earlier. The X-ray observation nearest in time to our {\sl HST} observation occurred on 2010 September 24 ({\sl XMM} observation 0605852201) and had a poorly constrained flux, consistent with the July 23 observation, albeit with lower signal-to-noise. Our  limit in F110W corresponds (e.g., \citealt{1979PASP...91..589B}) to a spectral flux limit $f_{\rm \nu, NIR} <  4.4\times10^{-31}$\,erg\,s$^{-1}$cm$^{-2}$Hz$^{-1}$, where we have assumed the reddening $A_V=0.7$ from the X-ray extinction measure \citep{1995A&A...293..889P,1998ApJ...500..525S}. Thus we derive a limit on the X-ray/NIR\footnote{Here we adopt $F=\nu f_\nu$ as a measure of the NIR/optical flux. The actual flux in a given filter, $F=\int R_\nu f_\nu d\nu\approx\Delta\nu f_\nu$, where $R_\nu$ is the filter throughput at frequency $\nu$, is smaller by a factor of a few. The definition  $F=\nu f_\nu$ is, however, convenient to compare fluxes measured with different broad filters.} flux ratio
   $F_{\rm X}/F_{\rm NIR} > 1000$. Likewise in the optical $f_{\rm \nu, O} <  2.3\times10^{-31}$\,erg\,s$^{-1}$cm$^{-2}$Hz$^{-1}$ and the X-ray/optical flux ratio is $F_{\rm X}/ F_{\rm  O} > 1100$. The flux ratio limits are still  consistent with the ratios found for persistent magnetars detected in the optical/NIR (see \citealt{2005ApJ...627..376D,2008A&ARv..15..225M}). We also retrieved the X-ray observation of \sgr\ closest in time {\em after} our {\sl HST} observation, by {\sl Chandra} ACIS (ObsID 13148 on 2010 November 29). We measure a flux of $(1.8\pm0.3)\times10^{-14}$\,erg\,s$^{-1}$cm$^{-2}$, indicating that the source had continued to fade. We plot the spectrum in Figure \ref{specs} alongside the one from 2010 July 23 \citep{2010Sci...330..944R}. The spectral peak shifted to  lower energies, indicating  a lower temperature (assuming a thermal spectrum).
   
Only three magnetars have so far been detected in the optical, 4U 0142+61 (a persistent AXP-type magnetar; \citealt{2000Natur.408..689H}), SGR 0501+4516 \citep{2008GCN..8160....1F} and 1E 1048.1$-$5937 (another AXP, detected only in the I-band; \citealt{2005ApJ...627..376D}). In each case, the observed emission is pulsed at the pulsar period, with pulsed fractions $>$50\%, i.e., higher than in soft X-rays \citep{2002Natur.417..527K,2011arXiv1106.1355D,2009MNRAS.394L.112D}. This clearly indicates the magnetospheric origin of the optical emission. It is illuminating to compare the optical properties of these magnetars with those of \sgr.

The measured magnitudes and inferred flux ratios of all the optically-detected magnetars are given in Table \ref{ratios} alongside our limits for \sgr. Our flux ratio limits are near the lowest values for the other magnetars, and imply that \sgr\ still could have similar spectral properties to the other magnetars, although the flux ratios must be higher than those for 1E 1048.1$-$5937. Note that the two AXPs have been observed in quiescence (but both are known to be variable), whereas both SGRs were observed after outbursts and may still have been fading. 

\begin{deluxetable}{cccccccccc}
\tablecaption{Optical-detected magnetars and their fluxes\label{ratios}}
\tabletypesize{\footnotesize}
\rotate
\tablewidth{0pt}
\tablehead{\colhead{Magnetar} & \colhead{$d$}&\colhead{$F_{\rm X}$}  & \colhead{NIR}  & \colhead{$\nu f_{\rm \nu, IR}$\tablenotemark{a}} & \colhead{Optical} & \colhead{$\nu f_{\rm \nu, O}$\tablenotemark{a}} & \colhead{$F_{\rm X}/F_{\rm NIR}$} & \colhead{$ F_{\rm X}/F_{\rm O}$} & \colhead{Refs\tablenotemark{b}}\\
 &\colhead{(kpc)}& \colhead{(0.5--10\,keV)}
}
\startdata 
4U 0142+61& $3.6\pm0.4$& 7.0$\times10^{-11}$ & $J=22.2$ & $1.2\times10^{-14}$& $V=25.6$& $2.9\times10^{-14}$&5700&2400 & 1\\
SGR 0501+4516\tablenotemark{c}  &  2.5$\pm$0.5&$8.7\times10^{-11}$& $K=19.2$ &$3.0\times10^{-14}$& $I=23.3$ & $6.9\times10^{-14}$& 2900&1300 & 2,3,4\\
 && $4\times10^{-11}$~\tablenotemark{d}& & & $i'=24.4$ & $4.2\times10^{-14}$& & 2100& 5\\
1E 1048.1$-$5937\tablenotemark{e} &9.0$\pm$1.7 &$7.5\times10^{-12}$ & $J=23.4$ & $7.4\times10^{-15}$ & $I=26.2$ & $7.2\times10^{-15}$& 1000&1000& 6\\
SGR 0418+5729 &2$\pm$0.5 &$1.2\times10^{-13}$ & $m_{\rm 110}>27.4$ & $<1.2\times10^{-16}$ & $m_{\rm 606}>28.6$ & $<1.1\times10^{-16}$& $>$1000& $>$1100& 7,8 
\enddata
\tablenotetext{a}{Here the NIR and optical fluxes are defined as $F=\nu f_\nu$for all the objects.}
\tablenotetext{b}{References: 1: \cite{2006ApJ...652..576D} 2: \citet{2008GCN..8159....1R} 3: \citet{2009ApJ...693L.122E} 4: \citet{2008GCN..8160....1F} 5: \cite{2011arXiv1106.1355D} 6: \citet{2005ApJ...627..376D} 7: \cite{2010Sci...330..944R} 8: this work.
}
\tablenotetext{c}{Here we list two X-ray and optical observations, shortly after outburst and when approaching quiescence}
\tablenotetext{d}{Near the quiescent value, and consistent with contemporaneous low-S/N {\sl SWIFT} observations (see text).}
\tablenotetext{e}{Values at quiescence.}
\tablecomments{The optical and infrared magnitudes listed are corrected for extinction before conversion to flux. The X-ray fluxes are unabsorbed. All fluxes are in units erg\,s$^{-1}$cm$^{-2}$. Distances are taken from \citet{2006ApJ...650.1070D} for the AXPs and from the locations of the Galactic spiral arms for the SGRs.}
\end{deluxetable}


The IR-X-ray spectra of the four magnetars in Table \ref{ratios} are shown in Figure \ref{specs}. The two AXPs  have similar multi-wavelength spectra, but the X-ray spectrum of SGR 0501+4516 appears to be closer to a power-law\footnote{Note that the extinction to SGR 0501+4516 is rather high and uncertain.} ($\Gamma\sim3$), whereas the spectrum of \sgr\ is more like a cooling black-body. 

The direction and low extinction to \sgr\ strongly suggests a distance of 2.0$\pm0.5$\,kpc, which is the distance to the Perseus arm (there is no further dense Galactic structure in this direction; \citealt{2002astro.ph..7156C}). For such a distance, the upper limits on NIR and optical luminosities ($L=4\pi d^2 \nu f_\nu$) are $L_{\rm F110W}<6\times10^{28}$\,erg\,s$^{-1}$, $L_{\rm F606W}<5\times10^{28}$\,erg\,s$^{-1}$, lower than the luminosities of any  detected magnetars.

The passive disk model for 4U 0142+61 by \citet{2006Natur.440..772W} predicts  $\nu f_\nu = 1.3\times10^{-14}$\,erg\,s$^{-1}$cm$^{-2}$ at 1.1\,$\mu$m, for the distance of 3.9\,kpc. If a similar disk were around \sgr, it would have  $\nu f_\nu =  5\times10^{-14}d^2_2 $\,erg\,s$^{-1}$cm$^{-2}$, where $d_2$ is the distance in units of 2\,kpc. We measure 
$\nu f_\nu < 1.2\times10^{-16}$\,erg\,s$^{-1}$cm$^{-2}$, i.e., 2.5 orders of magnitude smaller. For a disk  of the same size as in \citet{2006Natur.440..772W}, the disk inner temperature would need to be $T\leq600$\,K, i.e., cooler than plausible values for the (non-ice) dust sublimation temperature \citep{2011arXiv1104.5627K}. Alternatively, the disk could be more tenuous, or there may be no disk. Our measurements thus suggest that, if indeed disks contribute significantly to the luminosities of magnetars, then a different disk configuration (surface density and inner radius) is needed for \sgr\ compared to 4U 0142+61. If so, it could be connected to either the former being a transient magnetar and the latter persistent, or the much lower magnetic field of \sgr.

\citet{2011arXiv1102.0653M} discussed an alternative source of power for SGRs and AXPs: the spin-down of a rapidly rotating, magnetized WD. Our photomety limits allow us to place an upper limit on the temperature of a WD-sized ($R\sim10^9$\,cm) black-body emitter of $T\leq3000$\,K at 2\,kpc. Whereas WDs  with temperatures $T<4000$\,K are known (e.g., Durant et al. 2011, submitted), the cooling time required is $\tau \ga 10$\,Gyr. 
Our flux limits therefore constrain the WD interpretation for SGRs.

In conclusion, although \sgr\ is the nearest and least extincted magnetar known, we have not detected it in very deep optical and NIR observations. The source still offers a good opportunity to observe the optical spectrum of a magnetar, but this will only be possible following a new outburst. On the other hand, AXP 4U 0142+61 is the only persistent magnetar known whose optical/NIR spectrum can be measured, but the faintness requires extensive observation time.

\bigskip\noindent
Based on observations with {\sl HST}  (GO program 12183). This work is supported under grant HST-GO-12183.03-A by the Space Telescope Science Institute (STScI) and NASA grant NNX09AC84G.
The work by GGP was partly supported by the
Ministry of Education and Science of Russian Federation (Contract No.\ 11.G34.31.0001).

After this work appeared in preprint, we were contacted by J. Rueda, who pointed out that the model in Malheiro et al. (2011) specifically requires a high-mass white dwarf, with a radius, therefore, smaller than we assumed. Here we give the explicit dependence on radius of the limit one can place on the black-body surface temperature, one limit for each filter:
\begin{eqnarray}
T & < & \frac{12.48}{\ln(1+48.8 R_9^2)} \times 1000\,\rm{K}\\
T & < & \frac{24.39}{\ln(1+696 R_9^2)} \times 1000\,\rm{K}
\end{eqnarray}
where $R_9$ is the radius in units of 10,000\,km (10$^9$\,cm), for the F110W and F606W filters, respectively.

\bibliography{database}

\end{document}